# A Non Parametric Study of the Volatility of the Economy as a Country Risk Predictor

Sabatino Costanzo, Loren Trigo, Ramses Domínguez, William Moreno

*Abstract*

This paper intends to explain Venezuela's country spread behavior through the use of Neural Networks, incorporating as input data the IGAEM's (monthly economic activity general index, an index of economic indicators constructed by the BCV) VaR (value at risk), a measure of the shocks affecting country risk (country spreads) of emerging markets (the EMBI +Global) and EEUU's short term interest rates (6 month T-bills). The non parametric method used allows approaching the problem of finding the non linear relationship between these inputs and the country risk. The relative performance of the networks that were found was subsequently evaluated using the method of excess predictability (Anatolyev & Gerko, 2005). The network's performance in terms of predicting Venezuela's country risk behavior was satisfactory according to the evaluation method employed.

*Introduction*

The LEI's VaR, considered as a measure of the "economy's volatility", could be regarded as a complement to the traditional country-risk concept. This paper represents a continuation of the authors' research on the use of Neural Networks to forecast diverse phenomena in finance that the authors have been working on in the last several years. The country spread depends on the GDP (by construction) and the GDP on the LEI (by definition). The concept of risk is associated to the concept of volatility; therefore, the risk of a country not being capable of honoring its commitments (debt) is intuitively associated to the perceived and/or expected volatility of its economy. If we represent the economy by the GDP --or instead, by the LEI--, the size of "the risk of falling" associated to the volatility of either indicator (their VaR), should serve as an intuitively satisfactory way of measuring the risk a country has of not being capable of honoring its commitments, that is, an intuitively satisfactory default risk measure. (Notice that this idea is extensible to the measurement of credit quality of individual debtors).

On the other hand, if we consider that the debtor's default expectation depends not only on its economic stability but also on the interest rate imposed by its creditor, we can naturally infer the existence of a significant relationship between the global interest rates and the country spread, which is not linear.

Another advantage of non parametric methods over regressions in terms of the existing relationships between inputs and outputs, is that the respective roles of the inputs and outputs of two assets (a and b), are interchangeable. The effect that can immediately be deduced is that if input "a" in (t) can predict output "b" in (t + 1), then "b" in (t) –considered now as an input— can also predict output "a" in (t + 1), in other words, we can predict "a" and "b" in (t + 1) from "a" and "b" in (t). This "self-predictability" of portfolios throughout time, known as "cyclicity", is nothing but the ability that every dynamic system possesses to "evolve" throughout time.

This paper exploits, among other issues, the features of the relationship between interest rates of emerging nation's bonds issues and the changes in economic cycles studied by many such as, Erb, Harvey and Viskanta (1996); Neumeyer and Perri (2004), Uribe and Yue (2003), with the objective of finding an appropriate mechanism for predicting or estimating the behavior of Venezuela's country risk. As a referential variable of economic cycles, an economic activity follow-up index is used: the IGAEM (Monthly Economic Activity General Index, generated by the Venezuela's Central Bank), which is equivalent to the USA's Leading Economic Indicator (LEI), estimated by the Conference Board. This variable's effectiveness in predicting --or at least reflecting-- the economic cycles and the changes associated to the GDP, together with its intuitive nature and its accessibility, makes it an appropriate reference indicator for the national economy. Venezuela's Emerging Market Bonds Index Plus (EMBI+) was used as the objective variable (JP Morgan), also representing a widely used indicator as a country risk parameter, and given the methodology employed for its estimation, also as a general indicator associated to the interest rate movements of government securities.

In order to establish the prediction model, a non parametric method was used to solve the problem of finding the non linear relationships. We are specifically referring to neural networks, non linear models.

The result presents two dimensions: the development of the prediction model per se, and the important relationships found among the variables that were the subjects of this study. As for the model, ten neural networks with optimal performance were selected based on the Modified Sharpe

Index. They were subsequently engaged together in a master network responsible for the model's final estimation. The master network yielded a Modified Sharpe index of 17,14; furthermore, this network was subject to the Excess Predictability test of Anatolyev & Gerko, with a result of 99,86%, which can be considered highly satisfactory for a prediction tool.

The variables considered were the JP Morgan's EMBI+, the IGAEM, the EMBI+ Global and US Treasury bills' rates. The IGAEM's VaR predictive ability was confirmed, and it should improve with the support of the other two supplementary variables. Beyond that, this result makes evident the existence of a non-linear relationship between the variables involved, as it was previously ascertained.

The model built also shed some light as to which context is best for improving the prediction. A data set with the previously mentioned features allows estimating Venezuela's EMBI+ eight months in advance, even when it's necessary to wait until one month before the event for employing the master network, whose performance is substantially more dependable, even in terms of its consistency, on the medium term horizon. As a consequence of this search, the following question arises: If the IGAEM's Value at Risk (VaR) or LEI predicts Venezuela's EMBI+ in such an efficient manner, why not using it as an alternative indicator of the country risk? The method has interesting advantages given its intuitive nature and its estimation simplicity to estimate. Besides that, the VaR represents a volatility measure that could be related to the volatility of the Venezuelan economy.

The GDP is of special interest in terms of the government's ability to comply with its obligations and with those of the economy and to generate enough resources to improve public and private investments.

**Neural Networks**

Neural networks (back-percolation method) have been chosen to elaborate the estimations throughout this paper because, being non parametric statistical techniques (Smith, 1993), or non linear regression techniques (Sarle, 1998), is possible to extend its use to virtually every situation in which a relationship is presumed to exists between input variables (independent) and output variables (dependent), even where this relationship is complex or hard to articulate in the usual "correlation" or "differences between groups" terms.

From now on we will concentrate in four basic focal points that will be present throughout the whole paper:

1. The Emerging Markets Bond Index Plus (EMBI+); which has been used as Venezuela's country risk parameter and represents the specific "output" of the process.
2. The Value at Risk (VaR) of the so-called Monthly Economic Activity General Index (IGAEM)[1], also referred to Venezuela as an "input" of the process. (Issued by Venezuela's Central Bank on a monthly basis).
3. A *neural network* that works in an Excel environment.
4. A Modified Sharpe index, an especially interesting measure, since all the different networks that were yielded from the individual iterations are compared against it.

Complementarily, and due to the global economy's characteristics, two variables that support the proposed framework's theoretical robustness were incorporated: the JP Morgan EMBI+ Global and the U.S. Treasury bills' interest rates, both of which are also used as input variables in the process. All of the variables were expressed in their monthly form in order to comply with the restriction imposed by the IGAEM.

**Country Risk and the EMBI+**

*Growing globalization* is a frequently used phrase in specialized literature as an indication of the growing amount of investors that venture into alternative markets rather than their own in order to enjoy the benefits of international diversification in the case of capital markets; procuring expansions concerning demand, lowering their manpower costs, tax abatement, or simply lowering costs and maximizing the profitability of their own businesses. Such is the case of foreign direct investors.

In fact, world economic interdependency, revealed through the international movement of capitals, goods and services, has grown continually throughout the last five decades, Bouchet, Clarke and Groslambert (2003), Obstfeld and Taylor (2002). Complementing this long term fact, different studies show a "modest" tendency of investors to place capitals outside their local or domestic markets (French and Porteba (1995)), which according to Golup (1990), Tesar and Werner (1995), Pesenti and Wincoop (1996), is a phenomenon that occurs even in fixed income markets. This situation, although waning, is still present despite the known benefits of international diversification,

---

[1] In English: Leading Economic Indicator (LEI)

Grubel (1968), Grubel and Fadner (1971). One of the analyzed hypotheses in this context ascribes this *home bias* to the existence of information asymmetry handled by investors. Apparently, the comparative advantages of placing their capitals in local markets are *obvious* for them by virtue of the familiarity and knowledge they believe they have about the economic, political and social environment of their domestic areas. How can this bias be overcome? What kind of information could help investors earn an appropriate return for the assumed risk in their foreign investments?

The key for answering some of these questions may be the development of a pragmatic and intuitive methodology for the analysis of the so-called *country risk*. This "standardized indicator" intends to quantify the risks that could affect the financial performance of organizations and individual investors, assuming that these risks are the result of their exposure to specific economic, social and political conditions in foreign economies. Such conditions can be affected by factors of the most diverse nature: wars and social conflicts or xenophobia, changes on economic liberties or on fiscal policies, nationalization or asset freezing, foreign exchange controls and devaluations (Fiechter, Spillenkothen and Zamorski, 2002).

International Country Risk Guide, International Investor, Euromoney and rating providers such as Moody's, Standard and Poor's, Fitch and JP Morgan evaluate country risk with diverse quantification methods: some ponder various macroeconomic and financial variables, others use sociopolitical variables among which the are included: corruption, politics militarization, war on terrorism, etc., and still others whose quantification methodology is proprietary.

**Determining country risk through these agents is a complex process that implies studying qualitative factors, the creation of mathematical models and the evaluation of the investment's decision making process. An alternative for country risk measurement has been the use of interest rate differentials of bonds and risk free securities both in the primary and secondary markets. These make way to more clear and expedite methods that allow gathering the demands that the markets agents make in terms of return over a particular country's bond issues; methods which could very well be sustained by the Efficient Markets Hypothesis.**

In this last family of country risk estimators, the Emerging Markets Bond Index Plus should be included (EMBI+). This indicator of broad diffusion and ubiquitous use, has been generated by JP Morgan since 1994. It registers the total return (the result of capital gains and interest income) of foreign debt securities in emerging markets, expressing in a referential manner the *emerging debt* market's evolution and providing the researcher with a tool for studying the behavior of the debt of emerging countries.

The first elements selected for the index's construction are the countries and the financial securities. The nations selected, those corresponding to the emerging markets definition, include all the countries capable of paying their external debt and those whose credit ratings are at least BBB+/Baa12. In the general EMBI+, regional sub indexes are published as well as indexes by country. As of December 2000, this indicator included debt securities for 17 countries, even though nowadays, this number has increased to 27.

The securities that constitute the EMBI+ index have to be dollar denominated, have arrears lower than US$500 million, comply with certain liquidity requirements and are less than two years and a half away from its maturity date. The following debt securities are included for calculation purposes: Brady bonds, Eurobonds, dollar denominated debt securities issued in local markets and other loans. In addition, debt securities issued by these countries' private companies are incorporated, as long as they comply with the eligibility and liquidity criteria.

Throughout the index's estimation, a higher weight is assigned to the debt fluctuations of those countries whose importance is relatively higher within the total market of the selected countries. In order to achieve this, a relative weight is assigned to each debt issue while the remaining weight is assigned to the country taking into account the market capitalization.

**The EMBI+ represents credit and sovereign risk; in other words, a risk indicator that gathers the expectations of market agents on the probability that a government will default on its debt payment.**

In spite of this, many investors rely on this parameter for estimating their own returns and discount rates, regardless of the characteristics of their investments; this practice has propitiated an extended use of this risk indicator which, as we previously mentioned, has sovereign character. This fact is the source of the indicator's relevance, since it is well known that it has a direct incidence over financial institutions and companies when foreign investors rate them, Jüttner and McCarthy (2000); Hammer (2004); all this gives rise to the phrase used by Ferri (1999) in which he indicates that this sovereign risk is the *"pivot of all other country's ratings"*.

In our research we used the EMBI+ Global and EMBI+ Venezuela indexes. The EMBI+ Venezuela was constructed with the following securities:

**Chart**
**Source: Venezuelan Ministry of Finance (2003)**

| Venezuelan Bonds included in Country Risk indexes | | | | |
|---|---|---|---|---|
| | | Country Risk Index | | |
| Security | Outstanding amount in USD | EMBI+ | EMBI Global | EMBI Global Diversified |
| VE DCBs-DL | 1.040.818.075 | X | X | X |
| VE Brady Bonds Par - A | 1.638.400.000 | X | X | X |
| VE Brady Bonds Par – B | 719.500.000 | X | X | X |
| VE Republic 5 3/8% mat. 10 | 1.500.000.000 | X | X | X |
| VE Republic 10 3/4% mat. 13 | 1.487.389.000 | | X | X |
| VE Republic 13 5/8% mat. 18 | 500.000.000 | | X | X |
| VE Republic 7% mat. 18 | 1.000.000.000 | | X | X |
| VE Republic 9 1/4% mat. 27 | 4.000.000.000 | X | X | X |
| VE Republic 9 3/8% mat. 34 | 1.000.000.000 | | X | X |

**The IGAEM: Its Value at Risk**

As for the parameters used in this paper, a useful correlation that supports this study is the one found by Erb, Harvey and Viskanta (1996) between the country risk estimated by the composite-risk rating of the International Country Risk Guide and the fundamental valuation of the economic environment. Furthermore, Neumeyer and Perri (2004), Uribe and Yue (2003) add that fundamental shocks –-referring to changes in the different economic variables--, affect both the economic cycles and the country risk valuations in emerging markets. Beyond the debates on how to define and measure economic cycles, how to model them and how to predict behaviors and recession or growth tendencies –-as opposed to preceding studies--, our research used the IGAEM as a referential measure of economic cycles, seizing advantage of its predictability, Costanzo and Trigo (2004), which allows getting ahead of the changes experienced by the national economy.

The IGAEM is a version of the LEI or Composite Index of Leading Economic Indicators, which is simultaneously acknowledged as a convenient method for predicting business cycles. The LEI is a linear composite that has predicted each and every one of the eight recessions that have taken place in the United States. This index, published on a monthly basis by the Conference Board for the United States and Mexico, among others, is constructed with ten economic sub indexes. In Venezuela, its analog, the Monthly Economic Activity General Index (IGAEM) was designed by the Central Bank with the objective of relying on an indicator oriented towards evaluating the economy's evolution in the short run. In fact, the IGAEM allows analyzing national economic cycles as well as its changes, based on the battery of monthly indicators issued by the BCV (Central Bank of Venezuela). For the index's construction, a set of relevant indicators are employed in order to explain economic activity. Most of them are of physical production kind, and in the case of production amounts, are expressed in terms of volume through an appropriate conversion index. The IGAEM represents an average of the measured quantities, with fixed weights, in one base year. Its components are the primary indexes of physical production and demand related to sectoral activities of the Gross Domestic Product.

The set of selected indicators and the activities to which they are referred to are shown in Chart 1.

**Chart 1: IGAEM's components. Source: Central Bank of Venezuela (2000).**

| ACTIVITY | PARTIAL INDICATORS |
|---|---|
| OIL | Crude oil exports volume |
| | Derivatives exports volume |
| MINING | Iron production |
| PRIVATE MANUFACTURING | Private manufacturing industry production |
| POWER AND WATER SUPPLY | Generated power |
| CONSTRUCTION | Steal bar production |
| | Cement production |
| | Construction volume |
| | Public investments |
| COMMERCE | Commerce sales |
| FINANCIAL INSTITUTIONS AND INSURANCE COMPANIES | Value of checks in clearing houses |
| | Consumption through credit cards |
| SERVICES RENDERED FOR COMPANIES | Private manufacturing indexes |
| | Commerce indexes |
| IMPORTS RIGHTS | Imports indexes |

*Non Parametric Methods*

As previously mentioned, this study's problem is related to a data set with non linear characteristics, whereupon the use of neural networks has been suggested. This non linear estimation method has been used in solving many problems of economic and financial nature, and its growing use is a consequence of both the available computational power and the method's versatility and efficiency in finding behavioral patterns that result inaccessible to other mechanisms. The main advantage of neural networks consists in its great ability to detect and exploit the relationships between data, even in conditions where these are incomplete or in the presence of noise, which gives the network its *adaptive* character, in other words, it provides it with the capacity to learn from errors made in previous prediction attempts.

A neural network is a system made of a number of basic elements (Artificial Neurons) assembled in different layers, highly interconnected with each other (Synapses), that has both inputs and outputs, and that yields, through proper training, pertinent reactions to different entry stimuli. These systems imitate, to a certain degree, the human brain. Thereby, they require learning as well as a tutoring program and a tutoring routine to take over their training, based on historical knowledge of the problem.

The core of the performance of a neural network lies in a set of non linear functions which are evaluated in combinations of weighted stimuli that progressively move forward through the variables placed on each layer, until an exit flow is produced. The non linear map that links a layer's nodes with the nodes of other layers, does it through weight allocations, so that the weights assigned to each node, after processing by each non linear function, yield an arrangement of values that conveys towards each and every node of the next layer and so forth, until the last node is reached, the one corresponding to the network's exit. Paraphrasing one of Swanson and White's (1995) analogies, one could say that the input nodes constitute the door for stimuli that will be sent as signals to the next layer's nodes where they will be amplified or dampened according to their corresponding weights; immediately after that, a node from the hidden layer that receives these weighted signals will emit its own signal, correspondingly amplified or dampened. A basic illustration of the former explanation can be seen on Illustration 1.

**Illustration 1**

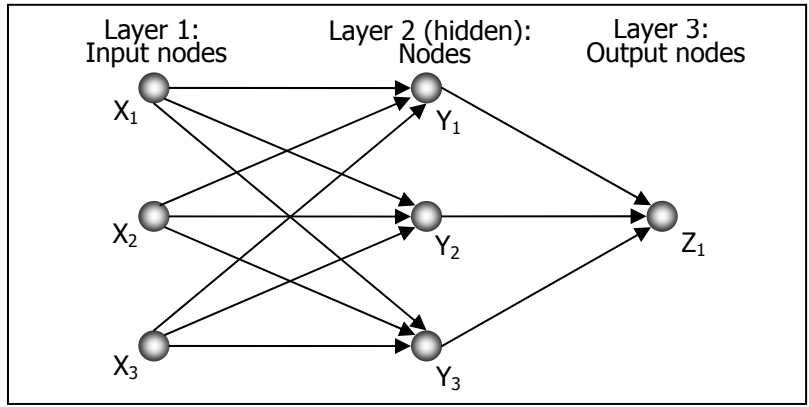

In other words, we can affirm that:

$Y_1 = f(w_1 X_1 + w_2 X_2 + w_3 X_3)$
$Y_2 = f(w_4 X_1 + w_5 X_2 + w_6 X_3)$
$Y_3 = f(w_7 X_1 + w_8 X_2 + w_9 X_3)$
$Z_1 = g(a_1 Y_1 + a_2 Y_2 + a_3 Y_3)$

In such expressions, coefficients $w_i$ y $a_i$ are called weights, while *f* y *g* represent the non linear functions. The network's training process will then consist of gradually executing successive *cycles* in which the weights are varied, ceteris paribus, until a satisfactory approximation is found, in terms of a provided parameter, to the specified exit. One of the most outstanding characteristics of this system is its adaptability to the network, which allows revisions of the approximation on each cycle, in the light of the errors made in previous cycles.

In the relevant literature we find different types of neural networks (Gately, 1996). The *feed forward multilayer network*, with a learning algorithm based on the *Back Propagation* technique (Rumelhart and McClelland, 1986), is the most popular network in economy and finance (Wong, 1995; Yao, Li and Tan, 1997), given the simplicity of the non linear functions involved. Despite this fact, we have favored the use of a more recent mechanism relying on the principle of *Back Percolation*, more efficient than *Back Propagation* because it seizes the advantages of a powerful algorithm of multiple simultaneous error reduction that eases the generation process of the experts.

The Back Percolation method assigns, for each cycle, sequential weight allocations to the nodes, all the way to the exits; this process occurs without evaluating the errors made during the cycle's execution. Back Percolation relies on a rebalancing mechanism of the input cell's weights, observing on each cycle the effect of such rebalancing over the network's exits. And so, during the cycle's execution, new weights that lower the existing difference between the network's predicted value and the observed real value are found.

In reference, for example, to Illustration 1, new weights that lower the error between the estimated Z and the observed Z will be determined. These new weights ($y_1$, $y_2$, $y_3$) will yield outputs that will lodge themselves in the network's hidden layer cells. In these layers, similar procedures to the one made on node Z will be applied, which will determine new weights, allowing the nodes or cells of this hidden layer to generate an output that complies with the function's exit requirements.

This method is efficient because all of the network's cells are simultaneously examined, improving their individual exit errors. Furthermore, the training process is much faster through this method.

The software used for processing neural networks was *Braincel*. This tool is capable of generating and evaluating different neural network architectures based on different combinations of possible parameters.

Using this application requires the following steps:

o   Manipulating and configuring data according to Braincel standards.
o   Defining data sets for training and testing.
o   Creating an expert file, which will include the parameter's structure and configuration for proper training.
o   Training the expert on the training data sets.
o   Testing the expert on the testing data sets.
o   Using the trained expert for making predictions.

The user should provide a data set so that the network trains itself in pattern searching, comparing its predictions with the historical ones. A second set (testing set) is used for supervising the network's training with the objective of eliminating the over adjustment. In general, it is recommended to use 60% of the data for training and 40% of the data for testing.

Braincel *self monitors* during the training process, controlling the internal configuration and the dynamics of the network. Some parameters can be manually configured during the expert's creation --the maximum acceptable error, extreme factors, learning rates and initial weights range, among others--.

**The Discriminant: Modified Sharpe Index**

A relationship derived from William F. Sharpe's proposed Sharpe Ratio, is the Sharpe Index. This index was used throughout this research as a discriminant and evaluation parameter of the network's performance. This index represents a risk adjusted performance measure for a specific investment portfolio. Its estimation presupposes establishing a relationship between the portfolio's excess over risk free returns and the portfolio's volatility, expressed in standard deviation terms.

The proposed modification, extracted from previous experiences regarding prediction on capital market instruments at IESA (2004-2005), combines three basic elements: The *Equity Curve* The *Perfect Equity Curve*, and the prediction's *average drawdown or average negative volatility*. The *Perfect Equity Curve,* also known as *Perfect Profit*, is achieved when 100% efficiency is obtained over an active investment strategy; that is, on each and every one of the occasions in which a long or short position is assumed, the strategy turns out to be successful. For its construction, the accrued returns of the portfolio's movements are estimated, without considering the period's variation sign. On the other hand, the *Equity Curve* is constructed in a similar fashion, but representing the real result of the active investment strategy. Hence, the sum of the real returns does consider the sign of the variation in terms of the assumed position. Therefore, a long position accompanied by a negative return, will cause a decrease in the returns accrued up to the date. In the opposite case, a short position with an asset's positive return generates equally negative results. The *average negative volatility* is simply calculated as the sum of the returns in which the assumed strategy failed, divided by the number of events; in other words, when the movement's prediction is erroneous and the asset moves in the opposite direction as expected, the negative volatility's corresponding amount will increase in the same proportion as the real variation experienced by the asset. Afterwards, this amount will be divided between the total number of mistakes. The formulas used during this procedure are the following:

$$return_t = \ln\left(\frac{x_t}{x_{t-1}}\right)$$

$$pe_t = pe_{t-1} + |return_t| \times 100$$

$$eq_t = eq_{t-1} + return_t \times 100$$

The expression used for completing the index's estimation has the following form:

$$ISM = \frac{Q}{Ave\,Negative\,Vol}$$

In such expression, the term *ISM* will correspond to the Modified Sharpe Index. The average volatility is represented by the term, *Ave Negative Vol*. Also, the ratio $Q$ relates the slopes of the lines of best fit of the equity curve ($m_{eq}$) and the perfect equity curve ($m_{pe}$), respectively.

The *ISM* parameter has been typified as a Sharpe Index by virtue of its similarity with the classic measure, which ultimately specifies the relationship between a variable's or portfolio's performance measure and its volatility.

Two additional aspects of the ISM structuring process directly involve the estimation of the lines of best fit of curves *eq* and *pe*. The methodology consists in adapting to the curves *eq* and *pe*, a factor between 1 and 11 that progressively amplifies all of the observations; from the first one, with factor 1, until the last one which will be multiplied by coefficient 11.

In order to achieve this, the following equations were used:

$$EQM_i = EQ_i \times (1 + 10 \times \frac{i}{n}) \qquad PEM_i = PE_i \times (1 + 10 \times \frac{i}{n})$$

In this case, $EQM_i$ and $PEM_i$ are the modified values of the *Equity Curve* and *Perfect Equity*, respectively, while $EQ_i$ and $PE_i$ represent the original values for both curves and *n* is the total number of data entries.

The line of best fit for these curves is established throughout the minimum squares method.

An additional test used as a complement to the ISM, is called *Excess Predictability* (EP), proposed by Anatolyev and Gerko (2005), similar to the well known *Directional Accuracy* (DA) by Pesaran and Timmermann (1992). As with the ISM, this test is tied to the buy/sell strategy performance of a *virtual financial advisor*. Anatolyev and Gerko describe it as a test based on a normalized profitability adjustable to the position achieved by a buy/sell strategy, according to a specific standard. In addition, like its predecessor, the DA, the results of the EP test are asymptotically distributed in a standard normal distribution (0, 1). The initial steps for the test's construction are similar to those used for determining the perfect equity and the equity curve, ultimately yielding the following expression:

$$EP \equiv \frac{A_T - B_T}{\sqrt{\hat{V}_{EP}}} \xrightarrow{d} N(0,1)$$

**Complementary Parameters: EMBI+Global & six month T-Bills**

*The EMBI+ (and in general, this type of differentials), is not only subservient to domestic factors. It can also be affected by exogenous factors or external shocks, such as movements on international interest rates or due to contagion as a result of external financial crisis.*

When a financial crisis simultaneously occurs in different countries, there could be, among other, three possible explanations:

- A coincidence of non related events in the different countries
- A shared economic shock and
- A contagion between investors because of crises in similar countries.

In recent years, such phenomenon has been experienced in the form of the "Tequila effect" in Mexico (1994), the Asian contagion (1997), or the 1998 Russian moratorium. This way, when a crisis develops in an emergent country, portfolio managers start to liquidate their positions, not only in the country in crisis, but also in those nations with similar profiles and performances to the stressed ones, whereupon the fall overflows frontiers in some sort of *domino effect*.

For this reason, the EMBI+ Global was selected for this model among the input variables in order to cover most of the aspects that could affect the country risk's behavior; particularly in this case, the effect derived from the generalized movements within the emerging markets.

"Country spreads serve as a transmission mechanism of world interest rates, capable of amplifying or dampening the effect of world-interest-rate shocks on the domestic economy" (Uribe, Martin and Yue, 2003, p. 3-4).

*This statement leads us to consider the effect (measured as the spread) that changes in interest rates have over the most important international financial markets.*

For example, Kamin and Kleist (1999) argue that short term low interest rates in industrialized countries' economies increase the demand of riskier investments – such characteristic includes emerging markets --, in the search for higher returns for investment portfolios. The spreads are not only determined by factors such as solvency. The high levels of global liquidity can lead to a fall in these values, until levels lower than the risk requirements accepted by many are achieved. Inversely, a rise in interest rates of more stable economies will generate a rise in financing throughout the world. As a consequence, a third input variable is included: the US treasury bills' returns as a reference of interest rates' movements in industrialized countries, also known as the quintessential "risk free" security.

## Data and Procedures

In order to establish network searching mechanisms capable of predicting the EMBI+ Venezuela, general methodological principles have allowed us to move forward until obtaining results that satisfy the generalized standards of predictive efficiency of the index's behavior. It is important to point out that the focal point of this analysis is the movement's direction prediction, not an approximation of the expected value of the objective's parameter. Nevertheless, the error's quantification is such that it weights the mistake's penalization in a proportional direction to the movement's magnitude that was erroneously predicted.

## The Data

The underlying data used in this study are the variables previously described: The IGAEM, the JP Morgan EMBI+ Global and the six month US T-Bills' interest rates, as the process's "input" variables, and the JP Morgan EMBI+ Venezuela as an "output". As for the IGAEM, data issued by Venezuela's Central Bank is available from January 1985 to April 2005 and issued on a monthly basis. All of the other data mentioned throughout this study –the JP Morgan EMBI+, both Venezuela's and the Global—and the six month US T-Bills' interest rates are provided by Reuters. All of the data has been expressed in monthly terms with the objective of suiting the IGAEM's periodicity. The dates of the relevant variables were adjusted, adapting them to the amount of the EMBI's available data, which, as we mentioned earlier, is the parameter with the lowest longevity in this study.

*Prediction Model Layout*

The general methodology is initiated with the linear regression of the relevant variables, as it is interesting to verify the existence of a linear relationship between the IGAEM and the EMBI+ Venezuela. If favorable, this procedure would allow to make predictions through linear projections in an expedite and robust way. For the available data, we found, after an extensive search, the following relation:

$\hat{y} = 1812,6 - 9,55(X - \bar{X})$, which is the linear equation for the IGAEM - EMBI+ case, and the set of results shown in Chart 2:

**Chart 2: Linear Regression Results**

| Regression's Statistics | |
|---|---:|
| Coefficient | -9.553796992 |
| Interception | 1812.603769 |
| Multiple correlation coefficient | 0.300004154 |
| R^2 determination coefficient | 0.090002492 |
| Probability | 0.004285703 |
| Observations | 89 |

This data reveals the existence of a weak linear relationship between the relevant variables, supported by a probability value for the null hypothesis (coefficient equivalent to cero) rejection of 0,43% significance. However, even though the linear relationship does exist to some extent, the regression model that was selected doesn't explain it adequately, as confirmed by an approximate $R^2$ of nine percent, which lead us to search for an alternative model.

The prediction then, gets to be explored through a non parametric method. In the case of a Neural Net, a sequence of steps for the network's actual training has been proposed. These previous steps respond to widespread consensus on the fact that, in order to take the most advantage of the underlying structural performance of neural networks, it is essential to preprocess the data.

***Data Preprocessing and Base Sets***

As we mentioned earlier, we do not recommend supplying to the network data, sets that haven't gone through the previous steps of preprocessing. The basic idea behind these strategies consists in eliminating the *chaotic noise* naturally immersed in the data that is regularly presented to the network, in particular, those that exhibit short term periodicity. One of these *preprocessing* procedures consist in smoothing the data in order to avoid "distractions" of the neural network in training, from the fundamental objective of deciphering the relevant patterns that contribute to the prediction. To such end, input data as well as output data have been subject to smoothing --as seen further along--, following the standard procedure for finding the networks that best predict the objective variable. For this purpose, ten *base sets* were used, each one of them yielding (due to the lags) "experts" or networks with independent and well differentiated responses and Sharpe indexes. These *base sets* correspond to a higher-order classification in which they were discriminated by type and/or amount of input and output data.

**Because the main objective of this study was to draw upon the movements of economic cycles (through the IGAEM's VaR) in order to predict movements in country risk, most of the manipulation or preprocessing done fell upon this parameter.**

Therefore, many of the sets show simple alterations on this variable and even its intentional omission within the input data, besides other minor effects of preprocessing for *complementary* and output data.

***Estimation of the IGAEM's volatility (VaR) - Historical Method***

Along with the IGAEM, its standard deviation was used, which could be conceptually equated with a non linear form of VaR. The reason for this is that even though VaR, in its strict definition, is a multiple of a negotiable asset's standard deviation (in which the factor is the size of the market position assumed), and even though the IGAEM isn't tradable or negotiable (and therefore lacks, strictly speaking, a VaR that recounts in absolute monetary terms the total amount subject to loss), its standard deviation is certainly the parameter with the most influence over the success of failure (winnings or losses) of the investments made in one nation as well as on the ability or inability of such nation to comply with its credit obligations.

In fact, the IGAEM's structure indicates, in relative terms, the variations in the economic cycles from a base year on. Consequently, in this context, the IGAEM's VaR would simply correspond to a quantification of the maximum negative percent variations of the index.

For the IGAEM's VaR estimation, the historical VaR technique was used, based on the idea of dividing the historical percent variations data in percentiles, and then searching for the datum *over which* X percent of the data is found, where *X* represents the confidence percent with which the VaR is supposed to be determined. An important aspect of the concept that was previously described is the phrase *over which*, because the objective is to delimit the data set that represents the largest losses, or in this case, the largest falls of the parameter being studied throughout the studied period. The obtained chain of value is unique once the ideal data period is established for the VaR's estimation.

In order to achieve the objectives, the study was divided in three stages:

o First, the IGAEM's monthly variations should be determined. The following formula was used for this purpose, with which the datum's geometrical variation was determined:

$$\%VARIATION = Ln\ (M)$$

Where M is the quotient of month t over month t-1. In this case, %VARIATION represents the variation taken place in month *t*. *Month $_n$* and *month $_{n-1}$* are the IGAEM's data that represent months *t* and *t-1*, respectively. Later on, this data were processed in order to convert them into the basic point system (bp), which allows establishing the equivalence between one percentage point and one hundred basic points (one percent is equivalent to 100 bps). The objective of this maneuver is to establish a better correspondence between the EMBI's data and the performance of the bonds used in this system. In addition, the positive component of the VaR was employed, that is, once the falls that corresponded to negative values were found, the appropriate sign changes were effectuated.

o Afterwards, the VaR's confidence percent was established. For this purpose, it was decided to work with 95 percent confidence, which is equivalent to a 95$^{th}$ percentile of data organized in ascending order or the lowest third based on the amount of data being used, as we'll explain on the next bullet. The 95 percent confidence represents the highest level of practical use of this parameter.

o Finally, the number of observations over which the VaR could be estimated is determined. With all of this in mind, sequential tests with different periods were effectuated and the results were compared through the *mean error* of the obtained VaR with the real variations. The expression that was used for this error was:

$$EAM = \frac{1}{n}\sum_{i=1}^{n}\left|estimated\ value_i - real\ value_i\right|$$

This formula, as well as the number of unusual points (outliers) found through *backtesting* (see Graph 2), allowed selecting the amount of data to be considered during the VaR's estimation. Such selection was effectuated over a period range that comprises 50 to 80 variation observations of the IGAEM. All of this data, mean error and outliers, were placed in matrixes that subsequently allowed, through an ocular inspection, determining which period yielded the best results.

**Chart 3: Result for the preprocessed VaR**

| Concept | Datum |
|---|---|
| *Lags for VaR estimation (months)* | 65 |
| *Confidence percent* | 95% |
| *Number of Outliers in 89 pieces of data* | 5 |
| *Mean Error* | 130,49 |

It is important to emphasize that according to the tests, both show dissimilar tendencies; that is, we found that with a high number of outliers, generally the mean error decreases and vice versa; it is not a constant relationship, however it is noticeable.

**Graph 2: Backtesting for the IGAEM's VaR (Non-preprocessed)**

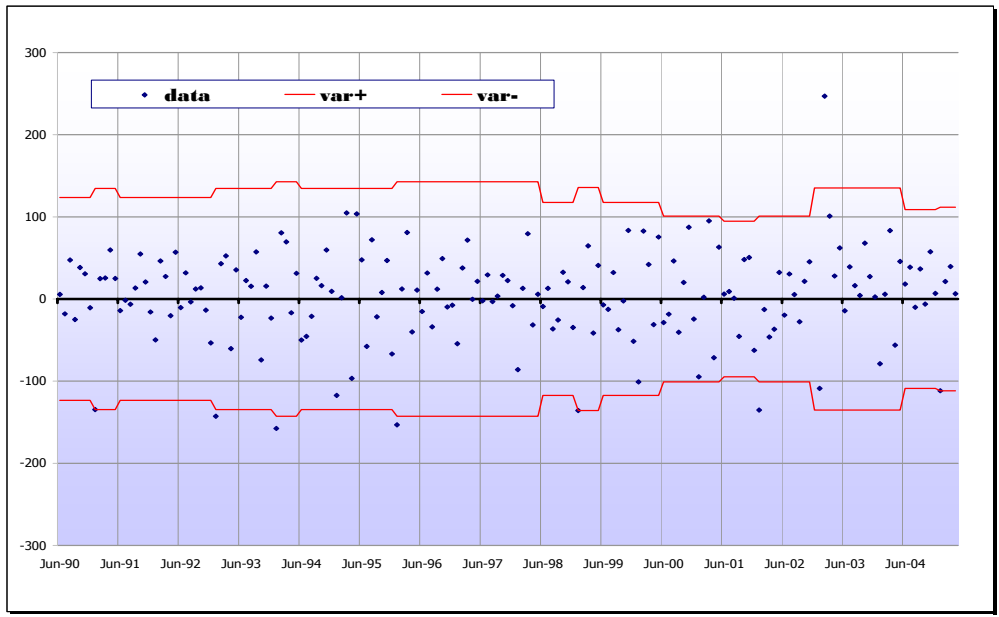

*Smooth Historical-IGAEM VaR Extension*

In terms of preprocessing the data, the VaR was smoothed. Its smoothened version substituted the historical measure that was discussed before. In this case, the smoothening method employed consisted of an exponential compound moving average, which allowed the creation of a data set that in many occasions superseded the VaR's originals in the training process of the neural network.

This method substitutes each datum for an average of the precedents. Therefore, the new value is a consequence of the real datum of the current period and of the preceding averages. Unlike the simple moving averages in which the mean is calculated based on a fixed number of data or *rectangular window* that moves longitudinally, the *compound moving averages* establish a triangular window that attenuates outlying data in time and makes emphasis on those of recent occurrence. In order to create this data sequence, the compound moving averages method relies on the following formula:

$$MA_t = \beta P_t + (1-\beta)MA_{t-1}$$

where beta is a number between 0 and 1. The moving average in month *t* is represented by $MA_t$ and the real datum is represented by $P_t$. Finally, $MA_{t-1}$ represents the previous moving average. As we explained earlier on, weighting the data sequence that intervenes in the average's estimation in ascending order is a natural consequence of the method. This degree of weightage is, at the same time, a function of the chosen *beta*. Thus, the closer to zero the value of *beta* is, the more weight will the most recent values have in the average's estimation. On the other hand, when *beta* is 1, $MA_t$ matches perfectly with $P_t$, real value, leaving the smoothening with no effect (case of the non smoothened VaR).

In a similar fashion to the previous situation, the determination of the optimal period for the VaR's estimation has been attempted, arguing that on this occasion an equivalent procedure is required for determining an appropriate beta. The resulting matrix of such process allowed concluding that the pair of data yielding the lowest mean error is the one with $\beta = 0,1$ and a lag of 65 data. The general result for this process was:

**Chart 4: Result for the preprocessed VaR**

| Concept | Datum |
|---|---|
| *Lags for estimating VaR* | 65 |
| B *value* | 0,1 |
| *Confidence percent* | 95% |
| *Number of Outliers in 89 pieces of data* | 56 |
| *Mean Error* | 48,23 |

An important characteristic of this modification is related to the results in terms of the outliers or unusual observations. A simple calculation would indicate the existence of 63 percent of values that go beyond the range of the established bands by the VaR, that is, in either case it would act as a referential value in equating the actual data with those obtained in the previous case.

**Graph 3: Backtesting for the IGAEM's VaR (Preprocessed)**

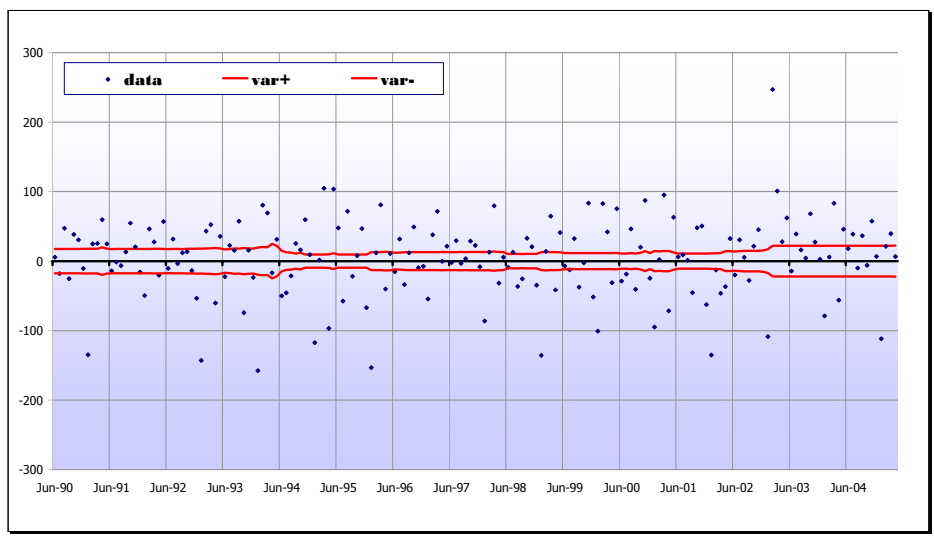

**Lags**

One of the much-employed procedures throughout the project was the creation of data subsets by main division or base set, which differ with each other only in the EMBI+'s month they were trying to predict. This was done with the objective of searching for cyclical patterns in the variables' behavior. For this purpose, it was decided to experiment with ten different lags between the input data dates and those of the output data. In consequence, the windows of fixed temporary length, 89 months, created over the input data, were deliberately lagged in terms of the date of the output data, one month at a time; so the tenth matrix constructed contains, for example, the input data that corresponds to February, aligned with the output data corresponding to the month of November. Each and every one of these discharges yielded a new matrix which has been named, in general, "Lag". These independent data were employed later on as training sets in order to generate, at the same time, ten different neural networks.

### Moving Average (MA) Over the IGAEM's VaR

In the *Smooth Historical-IGAEM VaR Extension* Section, we commented on one of the treatments applied to the data, the *Compound Moving Average*. That modification took place on multiple occasions over the VaR's data. An initial application yielded a data sequence that, as was mentioned earlier on, was alternated with the VaR per se, as an input variable in the experts' creation. A second application of this method, which has been identified as *level one* along with successive changes in the parameters and the addition of another *level (2)* of moving averages, which will be described in detail further along, allowed constructing various base sets based on variants of the same datum: the IGAEM's VaR.

**The result obtained through this process was the creation of matrixes of 15 to 18 input columns that when processed allowed proving the basic idea of the predictive potential that this version of the IGAEM has, as well as the economic cycles over the objective variable: Venezuela's EMBI+.**

### Introducing Block Averages (BA)

This preprocessing system which is similar to the *Compound Moving Average*, consists in averaging a window with *M+1* observations whose middle point occurred *n* dates before period *t*. *M* and *n* can adopt different values, however, the process generates some restrictions over their limits which will be subservient to the total number of samples being studied.

The expression that is used when applying Block Averages is the following:

$$BA(t-n) = \frac{1}{M+1} \sum_{k=-M/2}^{k=M/2} P(t-n+k)$$

**Preprocessing the EMBI+ Venezuela**

The last preprocessing effectuated maximized the smoothening of the output; in this case, the country risk or the EMBI+ Venezuela. Despite being simple, this method is very effective, because smoothing the output's movements gives the training network a better prediction capacity of the trajectory that the objective variable follows.

The two applications of this method were effectuated over the eight and ninth networks. The only difference between the two of them was how the VaR was used, with and without smoothening. Moreover, both networks counted with the presence of the complementary data. Subsequently, the method describing the last network created was employed (the tenth), --which also relied on the complementary data--, with the historical observations of the objective variable per se.

The method mentioned is defined by Ruggiero (2000) as a *normalized difference*, by the percent difference of the average of the last three real samples. In general terms, the procedure consisted of de-trending[2] and output-data-normalizing, which allows the mapping of events accumulated for long periods of time, that is, to detect if events that occurred 10 years ago are similar, in relative terms, to events that occurred yesterday. The formula used for this purpose is the following:

$$OUT_{mod} = \frac{OUT_t - OUT_{t-1}}{Mean(3days)}$$

in which $OUT_{mod}$ represents the modified output or the new output to be predicted; $OUT_t$ and $OUT_{t-1}$ indicate the real observations of *t* and *t-1*, respectively, while $Mean(3days)$ is the average of the real results that correspond to the previous three dates.

The methodology used in this case was the following: output modification, neural network training based on the modified output, new output prediction and finally, changes have to be reverted making the conversions that correspond to the previous expression.

$$OUT_{test} = OUT_{test\,mod} \times Mean(3days) + OUT_{t-1}$$

---

[2] Detrending, known as analysis and tendency suppression in time series' variables.

### Training Matrices Constitution

Once these steps are completed, data matrixes that intervened in the neural networks' training were created. In short, ten groups or base sets were created, most of them with ten subsets, made up of their own lags. Chart 5 displays a detailed composition of each base set.

**Chart 5: Base Sets Description**

| Base Set | Input Data | Output Data | Other Characteristics |
|---|---|---|---|
| 1 | *Non Smoothed IGAEM's VaR* <br> *EMBI+ Global* <br> *6 month US T-Bills* | *EMBI+ VENEZUELA* | PRACTICED OVER 10 LAGS |
| 2 | *Smoothed IGAEM's VaR* <br> *EMBI+ Global* <br> *6 month US T-Bills* | *EMBI+ VENEZUELA* | PRACTICED OVER 10 LAGS |
| 3 | *Double Smoothed IGAEM's VaR (both beta=0.1)* <br> *Moving Averages of the IGAEM's VaR* | *EMBI+ VENEZUELA* | ONLY ONE LAG |
| 4 | *Simple Smoothed IGAEM's VaR (beta=0.1)* <br> *Moving Averages of the IGAEM's VaR* | *EMBI+ VENEZUELA* | ONLY ONE LAG |
| 5 | *Double Smoothed IGAEM's VaR (ambos beta=0.1)* <br> *Moving Average of the IGAEM's VaR* <br> *EMBI+ Global* <br> *6 month US T-Bills* | *EMBI+ VENEZUELA* | ONLY ONE LAG |
| 6 | *Simple Smoothed IGAEM's VaR (both beta=0.1)* <br> *Moving Average of the IGAEM's VaR* <br> *EMBI+ Global* <br> *6 month US T-Bills* | *EMBI+ VENEZUELA* | ONLY ONE LAG |
| 7 | *EMBI+ Global* <br> *6 month US T-Bills* | *EMBI+ VENEZUELA* | PRACTICED OVER 10 LAGS |
| 8 | *Non Smoothed IGAEM's VaR* <br> *EMBI+ Global* <br> *6 month US T-Bills* | *EMBI+ VENEZUELA SMOOTHED* | PRACTICED OVER 10 LAGS |
| 9 | *Smoothed IGAEM's VaR* <br> *EMBI+ Global* <br> *6 month US T-Bills* | *EMBI+ VENEZUELA SMOOTHED* | PRACTICED OVER 10 LAGS |

| | | |
|---|---|---|
| 10 | *Smoothed IGAEM's VaR*<br>*EMBI+ Global*<br>*6 month US T-Bills*<br>*Previously Smoothed EMBI+ VZLA* | *EMBI+ PRACTICED VENEZUELA OVER 10 SMOOTHED LAGS* |

**Neural Network Training**

Finally, the neural networks were trained in order to create the experts. Each matrix was divided into two ranges: a training range, called *Training data* and the test range, which was denominated *Expanded testing data*. The proportion that was mostly used for each range was 60 and 40 percent of the whole data, respectively. Such proportion varied according to the amount of available data of the EMBI+ Venezuela, until reaching 45 percent, at the most, of the test range. Once the division had taken place, each *Training data* range was used for the individual training of an independent network throughout a *backpercolation* neural network. The task was programmed such that each network was trained 5000 times, randomly varying the initial weights that were assigned to each node. The steps *cycles* throughout the different layers were 1000 per iteration[3]. Each of the experts that were produced through this method was tested through its estimation, assigning as its entry a totally unknown data set (out-of-sample). These results were exposed to the scrutiny of the Modified Sharpe index and those that obtained the worst performance were systematically rejected until the best expert was chosen, that is, the best network among those found.

All of the other parameters established for the trainings were fixed according to Chart 6:

**Chart 6: Data for the network's re-training**

| Concept | Datum |
|---|---|
| *cycles* | 1.000 |
| *stop at error* | 10% |
| *initial learning rate* | 10% |
| *Randomized weights* | True |

---

[3] Iteration means each new training

***Using the Results in the Master Network's development***

The final step for the construction of the actual prediction model consisted in the creation of the so-called: *Master Network*. This process was completed based on the analysis of the experts obtained during the training phase and in the light of its results, in terms of the Modified Sharpe index. The best ten networks and their estimations were the base for creating the final matrix. The ulterior objective of this phase was finding a mechanism that allowed overcoming the divergences shown between the different networks that were obtained until that moment in terms of the upswing or downswing predictions of the EMBI+ Venezuela (Graph 4).

**Graph 4: Divergence Percentage in the Estimations by Date**

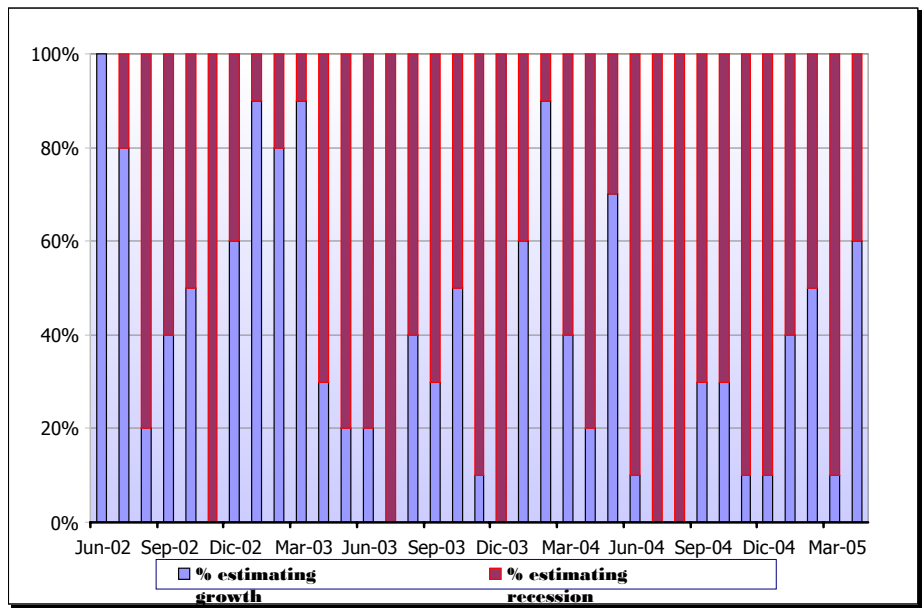

The percentages correspond to the number of networks forecasting increments in the EMBI+ Venezuela, against those forecasting its descent. Only the ten best networks are being considered.

As we mentioned earlier on, the master network's input data adequately correspond to the outputs of the best ten networks. On the other hand, the data was organized in a similar way as with the previous cases, with the obvious proviso that in this occasion no lag was introduced between the input and output data. We proceeded in consistency with the focal point of this study which is estimating the movement of the objective variable for the forthcoming month. Specifically, the input data for this test are the forthcoming month's estimations. Consequently, if the estimations made by the best networks correspond with month A, for example, these results ought to be introduced in order to predict—using the master network—the conclusive datum of the general model for month A.

In order to complete the master network, 5000 trainings were also executed. The same values that were used for the preceding networks were assigned to the rest of the Braincel processing parameters, and the processing included the discrimination throughout the Modified Sharpe index.

Finally, the suggested methodology for using this master network consists in collecting the estimations of other selected networks and using them as inputs when making the respective consult to the new network.

**Results Obtained**

The results obtained were very satisfactory. After laying out the test framework and its application to the creation of an adequate neural network for each case, ten high performance networks were obtained in terms of the prediction exactness of the EMBI+ Venezuela's behavior according to the established error parameters: the Modified Sharpe index and the *Excess Predictability* test proposed by Anatolyev and Gerko (2005). As we explained earlier, the proximity of the slopes of the *perfect equity* and *equity curve*, give faith of the prediction's quality. The result of many of the networks obtained in terms of the *equity curve* presented a slope angle that exceeded the line's slope bisector that best adapts itself to the *perfect equity* line, which is equivalent to saying that the relationship between the slopes is higher that ½, which can be considered as very satisfactory within the context of the estimations and predictions of economic and financial variables. The equity curve of the best networks, besides the one produced by the master network, are contained together with the perfect equity in Graph 5.

**Graph 5: Perfect Equity and the Best Equity Curves**

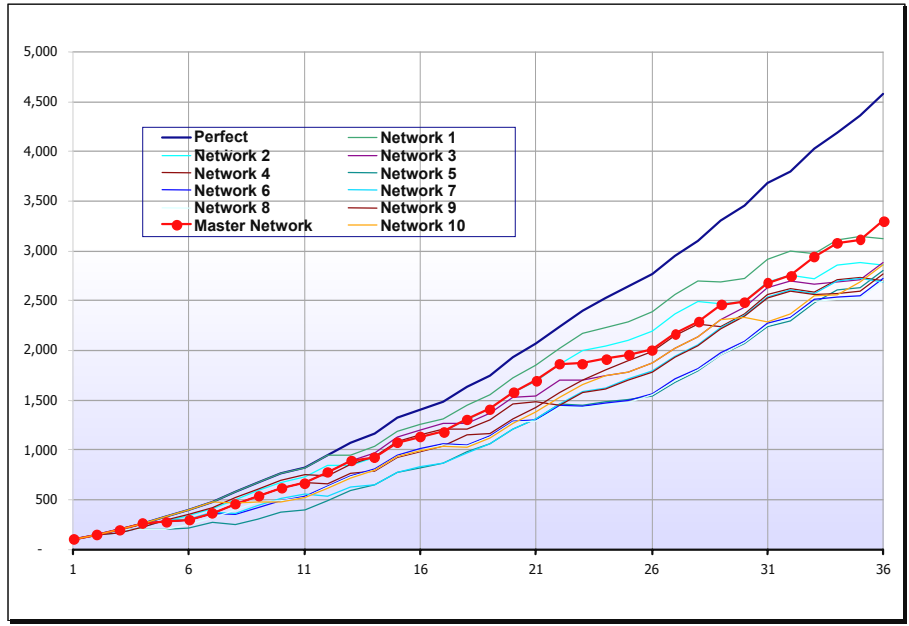

Note: The nomenclature used in this graph's caption is only referential, it does correspond itself with real name assigned to the networks, nor to the base sets.

The final result yielded by the master network was satisfactory. A Modified Sharpe index of 17,14 and a Norm EP of 99,86 percent for the master network confirms it. Even though in Graph 5 it is possible to ascertain that a pair of networks show higher results than those of the master network, in the initial months the negative volatility associated to the Master Network is much lower and the success of future predictions, seen through the projection of the final months of its equity curve, can be envisioned as with better perspectives. This statement is consistent with the construction method of the Sharpe index employed, which remarks the importance of the last predictions in terms of the ones that are temporarily a distance apart. In consequence, the line of best fit through the curve will have a higher slope that those that only showed better initial results, due to the depression that these last ones experience towards the end of the observation period. That is, the master network is, as expected, consistently better than the rest of the networks obtained.

Many facts stand out from the general results. These establish pronounced differences in terms of the predictive power of each and every one of the suggested models in the experts generating process, as well as with the specific characteristics of each of the variable sets employed.

The first facts are the ones that refer to the division made over the different *base sets*. For this analysis the base sets' results have been divided and their performances have been averaged.

Undoubtedly, ten samples per set are not sufficient in order to be able to establish conclusive positions, but it does let setting forth some notions about the preferences that could prevail when selecting one of these sets to predict the objective variable.

It should be pointed out that the best predictive answers for those sets with different Lags were those obtained from the base sets in which modifications in the output variable were performed (see Graph 6), that is, base sets *bestnet8*, *bestnet9* and *bestnet10*. However, the predictive potential presented by the sets constructed based on moving averages of the IGAEM's VaR can be also observed. The ability shown by the neural network for finding patterns over such sets generated an ISM of 7,62 and a mean Norm EP of 97 percent. In its first approximation, the achieved result (ISM: 5,24; mean Norm EP: 61,8 percent) with the set in which the IGAEM's VaR was deliberately omitted as input parameter (bestnet7) is also visible. The result by all means seems intuitive as it is expected that the market agents' demands over bond issues of a certain country respond, among other variables, to the local economy's movements; at least in emerging countries this is how it is expected to happen. That is, it is expected that a better approximation of the EMBI+ exists considering this reliance on, for example, *local* economic cycles, that taking only one variable set -- even when they are of interest to the estimation given its relationship--, are not a direct consequence of the economic, social and political performance of the nation.

**Graph 6: MOD Sharpe and Norm EP for the 10 Base Sets**

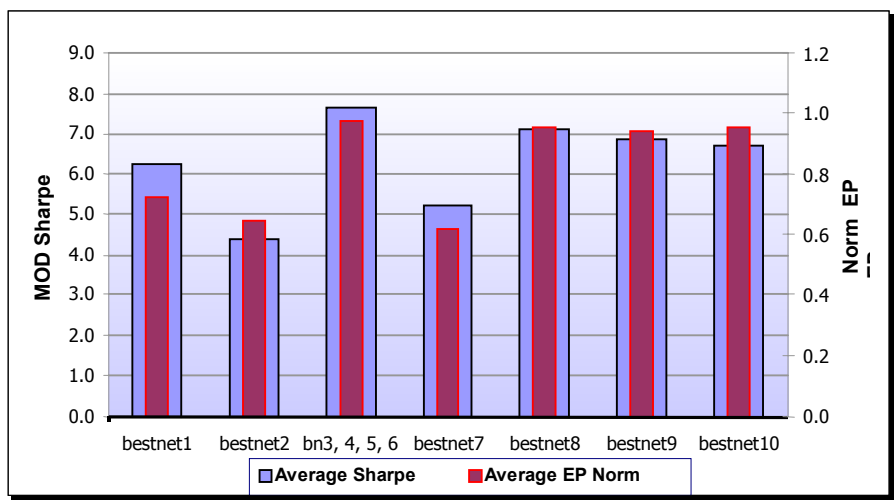

Another interesting aspect in the actual analysis was the usage and handling of the lags. In six out of the ten base sets this aspect was meticulously worked upon. The results (see Graph 7), highlights an important average performance for the eight months lag between the date in which the input conditions were generated and the date in which the EMBI+ Venezuela was produced. This statement is backed up by an SM of 7,166 and a Norm EP of 89,8 percent. Another interesting fact is that the next lag in terms of performance is the one that corresponds to nine months; that is, it could be inferred from this data that it is possible to foresee between eight to nine months in advance the events that will mark the directions that the EMBI+ Venezuela will follow, as variables such as the IGAEM's volatility, the EMBI+ Global and US six month T-Bills show their behavior.

**Graph 7: MOD Sharpe and Norm EB by Lag**

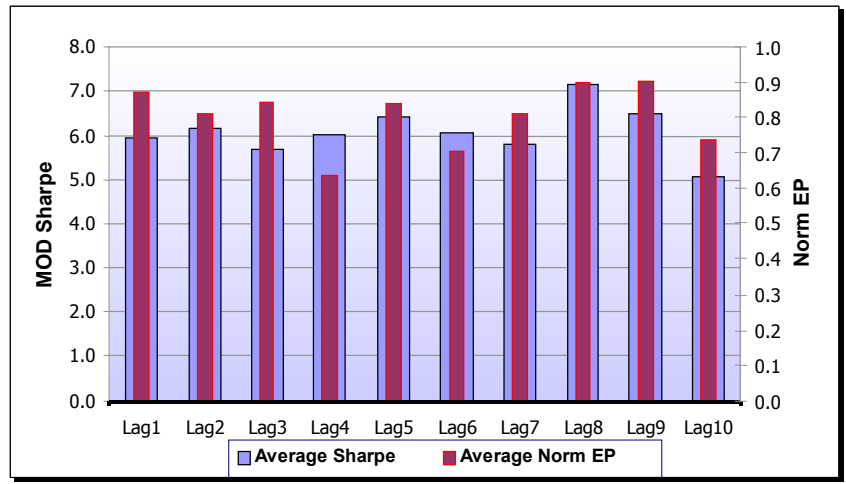

*Conclusion:*

**In short, as much as it seems reasonable to adopt the LEI as a substitute for the GDP when trying to find a representative index of a country's economy**
**—given that the LEI anticipates or coincides with the GDP while being a better representative of the economic activity and easier to predict--, it is as well as reasonable to argue in favor of adopting the LEI's VaR as a country risk substitute since the LEI's VaR as "volatility of the economy" fits not only as an intuitive measure of Country Risk, but is also a variable explaining Country Risk --when defined as EMBI-- through non parametric methods**
**—the relationship between LEI and EMBI is strong but not necessarily linear—This fact, together with the simplicity of their estimation method, qualify the VaR of the LEI as a true alternative definition for the level of risk of the Economy of a Country.**